\documentclass[12pt,prd,showpacs,tightenlines,nofootinbib]{revtex4}
\usepackage{bm}
\usepackage{graphics}
\usepackage{rotating}
\usepackage{epsfig}
\begin{document}
\title{\begin{flushright}{\rm\normalsize HU-EP-06/11}\end{flushright}
Semileptonic decays of heavy baryons in
the relativistic quark model}  
\author{D. Ebert}
\affiliation{Institut f\"ur Physik, Humboldt--Universit\"at zu Berlin,
Newtonstr. 15, D-12489  Berlin, Germany}
\author{R. N. Faustov}
\author{V. O. Galkin}
\affiliation{Institut f\"ur Physik, Humboldt--Universit\"at zu Berlin,
Newtonstr. 15, D-12489 Berlin, Germany}
\affiliation{Dorodnicyn Computing Centre, Russian Academy of Sciences,
  Vavilov Str. 40, 119991 Moscow, Russia}

\begin{abstract}

Semileptonic decays of heavy baryons consisting of one heavy
($Q=b,c$) and two light ($q=u,d,s$) quarks are considered in the
heavy-quark--light-diquark approximation. The relativistic
quasipotential equation is used for obtaining masses and wave
functions of both diquarks and baryons within the constituent
quark model. The weak transition matrix elements are expressed
through the overlap integrals of the baryon wave functions. 
The Isgur-Wise functions are determined in the whole accessible
kinematic range. The exclusive semileptonic decay rates and different
asymmetries are calculated with applying the heavy quark $1/m_Q$
expansion. The evaluated $\Lambda_b\to \Lambda_c l \nu$ decay 
rate agrees with its experimental value.

\end{abstract}

\pacs{13.30.Ce, 12.39.Ki, 14.20.Mr, 14.20.Lq}

\maketitle

\section{Introduction}
\label{intr}
The description of heavy baryon properties represents a very 
interesting and important
problem in quantum chromodynamics. Since the
baryon is a three-body system, its theory is much more complicated
compared to the two-body meson system.  The quark-diquark picture of a
baryon \cite{apefl,erv} is the popular approximation widely used to
describe the baryon properties \cite{apefl,erv,kkp,jaffe,wil}.  Such
approximation allows  to reduce the very complicated relativistic
three-body problem to the two-body one. Recently, we evaluated the
masses of the ground state heavy baryons in the 
framework of the relativistic quark model based on the quasipotential
approach \cite{efghbm}. The heavy-quark--light-diquark picture of the
heavy baryons was assumed. Both scalar and axial vector light diquarks
were considered. The relatively large size of the light
diquark was effectively taken into account by calculating the
diquark-gluon interaction form factor through the overlap integral of
the diquark wave functions. All the parameters
of the quark model had fixed values which were determined
from the previous studies of heavy and light meson
properties \cite{egf,efg,fg,lmm}. The overall reasonable agreement
(within a
few MeV) of our model predictions for heavy baryon masses with
the available experimental data supplies further support for the use
of the heavy-quark--light-diquark approximation. 

In this paper we continue the
study of heavy baryon properties and
apply our relativistic quark model for the calculation of their
exclusive semileptonic decays.  
Such investigations are important, since they provide an additional
source (complimentary to the heavy meson weak transitions)  for
determining the parameters of the Cabibbo-Kobayashi-Maskawa (CKM)
matrix, such as $V_{cb}$, from the comparison of the theoretical
predictions with the experimental data. We limit our present
consideration to the heavy-to-heavy ($b\to c$) transitions, where both
the initial and final baryons contain heavy quarks. For such
transitions the heavy quark effective theory (HQET), which is based on
the $1/m_Q$ expansion of the QCD Lagrangian and 
the emerging heavy quark
symmetry \cite{iwhs}, provides the most 
effective constraints on theoretical models and significantly reduces 
the number of independent form factors in each order of the heavy
quark expansion. In particular, transitions involving the $\Lambda_Q$
($Q=b,c$) 
baryons have the simplest structure, since the spectator light degrees
of freedom (the scalar diquark)
for these baryons have zero angular momentum. 
\footnote{The structure of the decay matrix elements for the
$\Lambda_Q$ baryons is simpler than for heavy mesons, since in the
latter case light degrees of freedom have spin 1/2.} 
In the heavy quark limit
only one universal form factor, the so-called Isgur-Wise function, is
required to describe the $\Lambda_Q\to \Lambda_{Q'}$ transition
\cite{iw,mrr}. At the subleading order of the heavy quark expansion one
mass parameter and one additional function emerge
\cite{ggw}. The consideration of the $\Omega_Q$ baryon  decays is
considerably more complicated, since light degrees of freedom 
(the axial vector diquark)
now have spin 1. 
It is necessary to introduce two functions for
parameterizing the $\Omega_Q\to 
\Omega^{(*)}_{Q'}$ transition in the heavy quark limit \cite{iw}, and five
additional functions and one mass parameter are needed at the
subleading order in $1/m_Q$ \cite{bb}. Note that transition matrix
elements between
baryons with spectator diquarks having different spins (e.g.,
$\Lambda_b\to \Sigma_c$, which violate isospin symmetry) vanish in the 
heavy quark limit and can proceed only due to the subleading 
corrections \cite{iw}. The heavy quark symmetry alone does not allow the
determination of the corresponding Isgur-Wise functions and mass
parameters. Only the 
normalization of some of these functions is known at
the point of zero recoil of the final heavy baryon. 
Thus, for 
the determination of these functions in the whole kinematic range
the application of  
nonperturbative methods is necessary.  Many different approaches were
previously used
for the calculation of the Isgur-Wise functions of
heavy baryons. 
However, most of them have important limitations. In
some of these approaches the Isgur-Wise functions are calculated  only
at one kinematic point or in a
limited region and then extrapolated to the whole kinematic range using 
an ad hoc ansatz, while other approaches
assume some parameterization for the heavy baryon
wave functions. The main aim of this paper is to determine
the corresponding
Isgur-Wise functions in the whole kinematic range through the overlap
integrals of the heavy baryon wave functions
in a consistent way within the relativistic quark model. 
These wave functions are known from the 
previous calculation of baryon masses \cite{efghbm}. 
On this basis exclusive
semileptonic decay rates and different asymmetries 
can then be evaluated
within the heavy quark expansion.

The paper is organized as follows. In Sec.~\ref{sec:rqm} we describe
our relativistic quark model and present predictions for the
masses of ground-state light diquarks and heavy baryons in the
heavy-quark--light diquark picture. In Sec.~\ref{mml} we discuss the
determination of the weak current matrix element between heavy baryon
states. The relativistic transformation of the baryon wave 
function from the rest to the moving reference frame is presented. The
general expressions for the weak matrix elements, decay rates and different
asymmetries for heavy baryons with scalar and axial vector diquarks
are given in Sec.~\ref{sec:ff}. In Sec.~\ref{sec:lld} semileptonic
decays of heavy baryons with the scalar diquark are
considered using the heavy quark expansion. Explicit expressions for the
leading and subleading Isgur-Wise functions are obtained as the
overlap integrals of the baryon wave functions. The predictions for
decay rates and the slope of the Isgur-Wise function are compared with
the experimental data for the $\Lambda_b\to 
\Lambda_c e \nu$ decay. Semileptonic decay rates of heavy baryons with
the axial vector diquark are studied in Sec.~\ref{sec:avd} in the
heavy quark limit. Finally, the comparison of our results for the
heavy baryon 
semileptonic decay rates with previous theoretical predictions and our
conclusions are given in Sec.~\ref{sec:dc}.

\section{Relativistic quark model for heavy baryons }
\label{sec:rqm}

In the quasipotential approach and quark-diquark picture of
heavy baryons \cite{efghbm} the interaction of two light quarks in a
diquark and the heavy 
quark interaction with a light diquark in a baryon are described by the
diquark wave function ($\Psi_{d}$) of the bound quark-quark state
and by the baryon wave function ($\Psi_{B}$) of the bound quark-diquark
state respectively,  which satisfy the
quasipotential equation  of the Schr\"odinger type 
\begin{equation}
\label{quas}
{\left(\frac{b^2(M)}{2\mu_{R}}-\frac{{\bf
p}^2}{2\mu_{R}}\right)\Psi_{d,B}({\bf p})} =\int\frac{d^3 q}{(2\pi)^3}
 V({\bf p,q};M)\Psi_{d,B}({\bf q}),
\end{equation}
where the relativistic reduced mass is
\begin{equation}
\mu_{R}=\frac{E_1E_2}{E_1+E_2}=\frac{M^4-(m^2_1-m^2_2)^2}{4M^3},
\end{equation}
and $E_1$, $E_2$ are given by
\begin{equation}
\label{ee}
E_1=\frac{M^2-m_2^2+m_1^2}{2M}, \quad E_2=\frac{M^2-m_1^2+m_2^2}{2M}.
\end{equation}
Here 
$M=E_1+E_2$ is the bound state mass (diquark or baryon),
$m_{1,2}$ are the masses of light quarks ($q_1$ and $q_2$) which form
the diquark or the masses
of the light diquark ($d$) and heavy quark ($Q$) which form
the heavy baryon ($B$), and ${\bf p}$  is their relative momentum.  
In the center of mass system the relative momentum squared on mass shell 
reads
\begin{equation}
{b^2(M) }
=\frac{[M^2-(m_1+m_2)^2][M^2-(m_1-m_2)^2]}{4M^2}.
\end{equation}

The kernel 
$V({\bf p,q};M)$ in Eq.~(\ref{quas}) is the quasipotential operator of
the quark-quark or quark-diquark interaction. It is constructed with
the help of the
off-mass-shell scattering amplitude, projected onto the positive
energy states. In the following analysis we closely follow the
similar construction of the quark-antiquark interaction in mesons
which were extensively studied in our relativistic quark model
\cite{egf}. For
the quark-quark interaction in a diquark we use the relation
$V_{qq}=V_{q\bar q}/2$ arising under the assumption about the octet
structure of the interaction  from the difference of the $qq$ and
$q\bar q$  colour states. An important role in this construction is
played by the Lorentz-structure of the confining  interaction. 
In our analysis of mesons while  
constructing the quasipotential of the quark-antiquark interaction, 
we adopted that the effective
interaction is the sum of the usual one-gluon exchange term with the mixture
of long-range vector and scalar linear confining potentials, where
the vector confining potential contains the Pauli terms.  
We use the same conventions for the construction of the quark-quark
and quark-diquark interactions in the baryon. The
quasipotential  is then defined by \cite{efghbm,efgm,egf} 

(a) for the quark-quark ($qq$) interaction
 \begin{equation}
\label{qpot}
V_{qq}({\bf p,q};M)=\bar{u}_{1}(p)\bar{u}_{2}(-p){\cal V}_{qq}({\bf p}, {\bf
q};M)u_{1}(q)u_{2}(-q),
\end{equation}
\[
{\cal V}_{qq}({\bf p,q};M)=\frac12\left[\frac43\alpha_sD_{ \mu\nu}({\bf
k})\gamma_1^{\mu}\gamma_2^{\nu}+ V^V_{\rm conf}({\bf k})
\Gamma_1^{\mu}({\bf k})\Gamma_{2;\mu}(-{\bf k})+
 V^S_{\rm conf}({\bf k})\right],
\]

(b) for quark-diquark ($Qd$) interaction
\begin{eqnarray}
\label{dpot} 
V_{Qd}({\bf p,q};M)&=&\frac{\langle d(P)|J^{\mu}|d(Q)\rangle}
{2\sqrt{E_d(p)E_d(q)}} \bar{u}_{Q}(p)  
\frac43\alpha_sD_{ \mu\nu}({\bf 
k})\gamma^{\nu}u_{Q}(q)\cr
&&+\psi^*_d(P)\bar u_Q(p)J_{d;\mu}\Gamma_Q^\mu({\bf k})
V_{\rm conf}^V({\bf k})u_{Q}(q)\psi_d(Q)\cr 
&&+\psi^*_d(P)
\bar{u}_{Q}(p)V^S_{\rm conf}({\bf k})u_{Q}(q)\psi_d(Q), 
\end{eqnarray}
where $\alpha_s$ is the QCD coupling constant;
 $\langle d(P)|J_{\mu}|d(Q)\rangle$ is the vertex of the 
diquark-gluon interaction which takes into account the diquark size
\cite{efghbm} in terms of the diquark wave function overlap  
$\Big[$$P=(E_d,-{\bf p})$ and $Q=(E_d,-{\bf q})$,
$E_d=(M^2-m_Q^2+M_d^2)/(2M)$ $\Big]$. 
$D_{\mu\nu}({\bf k})$ is the 
gluon propagator in the Coulomb gauge
\begin{equation}
D^{00}({\bf k})=-\frac{4\pi}{{\bf k}^2}, \quad D^{ij}({\bf k})=
-\frac{4\pi}{k^2}\left(\delta^{ij}-\frac{k^ik^j}{{\bf k}^2}\right),
\quad D^{0i}=D^{i0}=0,
\end{equation}
and ${\bf k=p-q}$; $\gamma_{\mu}$ and $u(p)$ are 
the Dirac matrices and spinors
\begin{equation}
\label{spinor}
u^\lambda({p})=\sqrt{\frac{\epsilon(p)+m}{2\epsilon(p)}}
\left(\begin{array}{c}
1\\ \displaystyle\frac{\mathstrut\bm{\sigma}{\bf p}}
{\mathstrut\epsilon(p)+m}
\end{array}\right)
\chi^\lambda,
\end{equation}
with $\epsilon(p)=\sqrt{{\bf p}^2+m^2}$. 

The diquark state in the confining part of the quark-diquark
quasipotential (\ref{dpot}) is described by the wave functions
\begin{equation}
  \label{eq:ps}
  \psi_d(p)=\left\{\begin{array}{ll}1 &\qquad \text{ for scalar diquark}\\
\varepsilon_d(p) &\qquad \text{ for axial vector diquark}
\end{array}\right. ,
\end{equation}
where the four vector
\begin{equation}\label{pv}
\varepsilon_d(p)=\left(\frac{(\bm{\varepsilon}_d {\bf
p})}{M_d},\bm{\varepsilon}_d+ \frac{(\bm{\varepsilon}_d {\bf p}){\bf
  p}}{M_d(E_d(p)+M_d)}\right)  
\end{equation} 
is the polarization vector [$\varepsilon^\mu_d(p) p_\mu=0$] of the
axial vector 
diquark with momentum ${\bf p}$, $E_d(p)=\sqrt{{\bf p}^2+M_d^2}$ and
$\varepsilon_d(0)=(0,\bm{\varepsilon}_d)$ is the polarization vector in
the diquark rest frame. The effective long-range vector vertex of the
diquark can be presented in the form  
\begin{equation}
  \label{eq:jc}
  J_{d;\mu}=\left\{\begin{array}{ll}
  \frac{\displaystyle (P+Q)_\mu}{\displaystyle
  2\sqrt{E_d(p)E_d(q)}}&\qquad \text{ for scalar diquark}\cr
-\frac{\displaystyle (P+Q)_\mu}{\displaystyle2\sqrt{E_d(p)E_d(q)}}
  +\frac{\displaystyle i\mu_d}{\displaystyle 2M_d}\Sigma_\mu^\nu 
\tilde k_\nu
  &\qquad \text{ for axial 
  vector diquark}\end{array}\right. ,
\end{equation}
where $\tilde k=(0,{\bf k})$. Here the antisymmetric tensor 
reads
\begin{equation}
  \label{eq:Sig}
  \left(\Sigma_{\rho\sigma}\right)_\mu^\nu=-i(g_{\mu\rho}\delta^\nu_\sigma
  -g_{\mu\sigma}\delta^\nu_\rho),
\end{equation}
and the spin ${\bf S}_d$ of the axial vector diquark is given by
$(S_{d;k})_{il}=-i\varepsilon_{kil}$. We choose the total
chromomagnetic moment of the axial vector 
diquark $\mu_d=0$ \cite{efgtetr}. Such a choice appears to be natural,
since the long-range chromomagnetic interaction of the 
diquark proportional to $\mu_d$ then  vanishes in accord with the
flux tube model. 

The effective long-range vector vertex of the quark is
defined by \cite{egf,sch}
\begin{equation}
\Gamma_{\mu}({\bf k})=\gamma_{\mu}+
\frac{i\kappa}{2m}\sigma_{\mu\nu}\tilde k^{\nu}, \qquad \tilde
k=(0,{\bf k}),
\end{equation}
where $\kappa$ is the Pauli interaction constant characterizing the
anomalous chromomagnetic moment of quarks. In the configuration space
the vector and scalar confining potentials in the nonrelativistic
limit reduce to
\begin{eqnarray}
V^V_{\rm conf}(r)&=&(1-\varepsilon)V_{\rm conf}(r),\nonumber\\
V^S_{\rm conf}(r)& =&\varepsilon V_{\rm conf}(r),
\end{eqnarray}
with 
\begin{equation}
V_{\rm conf}(r)=V^S_{\rm conf}(r)+
V^V_{\rm conf}(r)=Ar+B,
\end{equation}
where $\varepsilon$ is the mixing coefficient.

The constituent quark masses $m_b=4.88$ GeV, $m_c=1.55$ GeV,
$m_u=m_d=0.33$ GeV, $m_s=0.5$ GeV and 
the parameters of the linear potential $A=0.18$ GeV$^2$ and $B=-0.3$ GeV
have the usual values of quark models.  The value of the mixing
coefficient of vector and scalar confining potentials $\varepsilon=-1$
has been determined from the consideration of charmonium radiative
decays \cite{efg} and the heavy quark expansion \cite{fg}. 
Finally, the universal Pauli interaction constant $\kappa=-1$ has been
fixed from the analysis of the fine splitting of heavy quarkonia ${
}^3P_J$- states \cite{efg}.  Note that the 
long-range chromomagnetic contribution to the potential in our model
is proportional to $(1+\kappa)$ and thus vanishes for the 
chosen value of $\kappa=-1$.

The quasipotential (\ref{qpot}) can  be used for arbitrary quark
masses.  The substitution 
of the Dirac spinors  into (\ref{qpot}) results in an extremely
nonlocal potential in the configuration space. Clearly, it is very hard to 
deal with such potentials without any additional transformations.
 In oder to simplify the relativistic $q q$ potential, we make the
following replacement in the Dirac spinors \cite{efghbm,egf,lmm}:
\begin{equation}
  \label{eq:sub}
  \epsilon_{1,2}(p)=\sqrt{m_{1,2}^2+{\bf p}^2} \to E_{1,2}.
\end{equation}
This substitution makes the Fourier transformation of the potential
(\ref{qpot}) local, but the resulting relativistic potential becomes
dependent on the diquark and baryon masses in a very complicated
nonlinear way.  
We consider only the baryon ground
states, which further simplifies our analysis, since all terms
containing orbital momentum vanish. The detailed expressions for the
relativistic quark potential can be found in Ref.~\cite{efghbm}.
The obtained masses of the light diquarks are given in
Table~\ref{tab:mass}. The heavy baryon masses calculated in the
heavy-quark--light diquark approximation are presented in
Table~\ref{tab:bm}  
in comparison with the available experimental data \cite{pdg}. 
There an overall good agreement of our predictions with 
experiment is found.

\begin{table}
  \caption{Masses of light ground state diquarks (in MeV). S and A
    denote scalar and axial vector diquarks which are
antisymmetric $[q,q']$ and symmetric $\{q,q'\}$ in flavour
indices, respectively. }
  \label{tab:mass}
\begin{ruledtabular}
\begin{tabular}{ccc}
Quark content& Diquark type& Mass\\
\hline
$[u,d]$& S & 710 \\
$\{u,d\}$& A & 909  \\
$[u,s]$ & S& 948  \\
$\{u,s\}$& A & 1069 \\
$\{s,s\}$& A & 1203  
  \end{tabular}
\end{ruledtabular}
\end{table}

\begin{table}
\caption{\label{tab:bm} Masses of the ground state heavy baryons (in MeV).}
\begin{ruledtabular}
\begin{tabular}{cccc}
Baryon &$I(J^P)$& $M^{\rm theor}$ \cite{efghbm}
& $M^{\rm exp}$~\cite{pdg}\\
\hline
$\Lambda_c$&$0(\frac12^+)$ & 2297&2284.9(6)   \\
$\Sigma_c$&$1(\frac12^+)$ &2439 &2451.3(7)  \\
$\Sigma^*_c$&$1(\frac32^+)$ & 2518 &2515.9(2.4) \\
$\Xi_c$&$\frac12(\frac12^+)$& 2481 &2466.3(1.4)  \\
$\Xi'_c$&$\frac12(\frac12^+)$&2578 &2574.1(3.3)\\ 
$\Xi^*_c$&$\frac12(\frac32^+)$&2654 &2647.4(2.0)\\
$\Omega_c$ & $0(\frac12^+)$& 2698 &2697.5(2.6)\\
$\Omega^*_c$ & $0(\frac32^+)$&2768 &\\
$\Lambda_b$&$0(\frac12^+)$& 5622  & 5624(9)   \\
$\Sigma_b$&$1(\frac12^+)$ &5805 &  \\
$\Sigma^*_b$&$1(\frac32^+)$ & 5834&\\
$\Xi_b$&$\frac12(\frac12^+)$& 5812&\\
$\Xi'_b$&$\frac12(\frac12^+)$&5937&\\
$\Xi^*_b$&$\frac12(\frac32^+)$&5963&\\
$\Omega_b$ & $0(\frac12^+)$&6065&\\
$\Omega^*_b$ & $0(\frac32^+)$ & 6088& \\
\end{tabular}
\end{ruledtabular}
\end{table}

\section{Matrix elements of the weak current for heavy baryon
  decays} \label{mml}  

In order to calculate the exclusive semileptonic decay rate of the
heavy baryon, it is necessary to determine the corresponding matrix
element of the  weak current between baryon states.
In the quasipotential approach,  the matrix element of the weak current
$J^W_\mu=\bar Q'\gamma_\mu(1-\gamma_5)Q$, associated with the
heavy-to-heavy quark $Q\to Q'$ ($Q=b$ 
and $Q'=c$) transition, between baryon states with masses $M_{B_Q}$,
$M_{B_{Q'}}$ and momenta $p_{B_Q}$, $p_{B_{Q'}}$ takes the form \cite{f} 
\begin{equation}\label{mxet} 
\langle B_{Q'}(p_{B_{Q'}}) \vert J^W_\mu \vert B_Q(p_{B_Q})\rangle
=\int \frac{d^3p\, d^3q}{(2\pi )^6} \bar \Psi_{B_{Q'}\,{\bf p}_{B_{Q'}}}({\bf
p})\Gamma _\mu ({\bf p},{\bf q})\Psi_{B_Q\,{\bf p}_{B_Q}}({\bf q}),
\end{equation}
where $\Gamma _\mu ({\bf p},{\bf
q})$ is the two-particle vertex function and  
$\Psi_{B\,{\bf p}_B}$ are the
baryon ($B=B_Q,B_{Q'})$ wave functions projected onto the positive
energy states of 
quarks and boosted to the moving reference frame with momentum ${\bf p}_B$.
\begin{figure}
  \centering
%\vskip 1.5cm
  \includegraphics{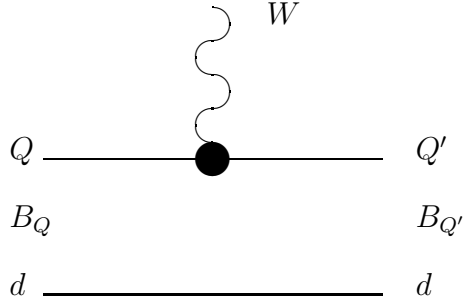}
\caption{Lowest order vertex function $\Gamma^{(1)}$
contributing to the current matrix element (\ref{mxet}). \label{d1}}
\end{figure}

\begin{figure}
  \centering
%\vskip 1.5cm
  \includegraphics{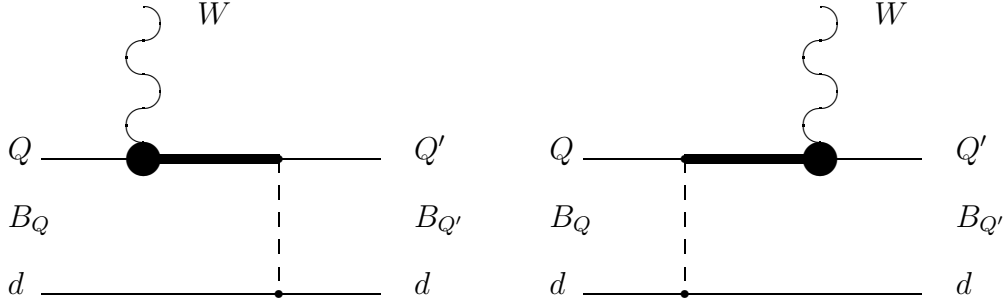}
\caption{ Vertex function $\Gamma^{(2)}$
taking the quark interaction into account. Dashed lines correspond  
to the effective potential ${\cal V}_{Qd}$ in 
(\ref{dpot}). Bold lines denote the negative-energy part of the quark
propagator. \label{d2}}
\end{figure}

 The contributions to $\Gamma$ come from Figs.~\ref{d1} and \ref{d2}. 
The contribution $\Gamma^{(2)}$ is the consequence
of the projection onto the positive-energy states. Note that the form
of the 
relativistic corrections resulting from the vertex function
$\Gamma^{(2)}$ is explicitly dependent on the Lorentz structure of the
quark-diquark interaction. In the heavy quark limit $m_{Q}\to \infty$ 
only $\Gamma^{(1)}$ contributes, while $\Gamma^{(2)}$  
gives the subleading order contributions. 
The vertex functions are given by
\begin{equation} \label{gamma1}
\Gamma_\mu^{(1)}({\bf
p},{\bf q})=\psi^*_d(p_d)\bar
u_{Q'}(p_{Q'})\gamma_\mu(1-\gamma^5)u_Q(q_Q)\psi_d(q_d) 
(2\pi)^3\delta({\bf p}_d-{\bf
q}_d),\end{equation}
and
\begin{eqnarray}\label{gamma2} 
\Gamma_\mu^{(2)}({\bf
p},{\bf q})&=& \psi^*_d(p_d)\bar u_{Q'}(p_{Q'}) \Bigl\{\gamma_{\mu}(1-\gamma^5)
\frac{\Lambda_Q^{(-)}(
k)}{\epsilon_Q(k)+\epsilon_Q(p_{Q'})}\gamma^0
{\cal V}_{Qd}({\bf p}_d-{\bf
q}_d)\nonumber \\ 
& &+{\cal V}_{Qd}({\bf p}_d-{\bf
q}_d)\frac{\Lambda_{Q'}^{(-)}(k')}{ \epsilon_{Q'}(k')+
\epsilon_{Q'}(q_Q)}\gamma^0 \gamma_{\mu}(1-\gamma^5)\Bigr\}u_Q(q_Q)
\psi_d(q_d),\end{eqnarray}
where the superscripts ``(1)" and ``(2)" correspond to Figs.~\ref{d1} and
\ref{d2},  ${\bf k}={\bf p}_{Q'}-{\bf\Delta};\
{\bf k}'={\bf q}_Q+{\bf\Delta};\ {\bf\Delta}=
M_{B_{Q'}}{\bf v'}-M_{B_Q}{\bf v}$; \ $\epsilon(p)=\sqrt{m^2+{\bf p}^2}$;
$$\Lambda^{(-)}(p)=\frac{\epsilon(p)-\bigl( m\gamma
^0+\gamma^0({\bm{ \gamma}{\bf p}})\bigr)}{ 2\epsilon (p)}.$$
Here \cite{f} 
\begin{eqnarray} 
p_{Q',d}&=&\epsilon_{Q',d}(p)v'
\pm\sum_{i=1}^3 n^{(i)}(v')p^i, \qquad v'^\mu=\frac{p^\mu_{B_{Q'}}}{M_{B_{Q'}}},\cr
q_{Q,d}&=&\epsilon_{Q,d}(q)v \pm \sum_{i=1}^3 n^{(i)}
(v)q^i, \qquad v^\mu=\frac{p^\mu_{B_Q}}{M_{B_Q}},\end{eqnarray}
and $n^{(i)}$ are three four-vectors given by
$$ n^{(i)\mu}(v)=\left\{ v^i,\ \delta_{ij}+
\frac{v^iv^j}{v^0+1}\right\}.$$

It is important to note that the wave functions entering the weak current
matrix element (\ref{mxet}) are not in the rest frame in general. For
example,   
in the $B_Q$ baryon rest frame (${\bf v}=0$), the final  baryon
is moving with the recoil momentum ${\bf \Delta}$. The wave function
of the moving  baryon $\Psi_{{B_{Q'}}\,{\bf\Delta}}$ is connected 
with the  wave function in the rest frame 
$\Psi_{{B_{Q'}}\,{\bf 0}}\equiv \Psi_{B_{Q'}}$ by the transformation \cite{f}
\begin{equation}
\label{wig}
\Psi_{{B_{Q'}}\,{\bf\Delta}}({\bf
p})=D_{Q'}^{1/2}(R_{L_{\bf\Delta}}^W)D_d^{\cal I}(R_{L_{
\bf\Delta}}^W)\Psi_{{B_{Q'}}\,{\bf 0}}({\bf p}),\qquad  {\cal I}=0,1,
\end{equation}
where $R^W$ is the Wigner rotation, $L_{\bf\Delta}$ is the Lorentz boost
from the baryon rest frame to a moving one, and   
the rotation matrix of the heavy quark spin $D^{1/2}(R)$ in spinor
representation is given by 
\begin{equation}\label{d12}
{1 \ \ \,0\choose 0 \ \ \,1}D^{1/2}_{Q'}(R^W_{L_{\bf\Delta}})=
S^{-1}({\bf p}_{Q'})S({\bf\Delta})S({\bf p}),
\end{equation}
where
$$
S({\bf p})=\sqrt{\frac{\epsilon(p)+m}{2m}}\left(1+\frac{\bm{\alpha}{\bf p}}
{\epsilon(p)+m}\right)
$$
is the usual Lorentz transformation matrix of the four-spinor. The
rotation matrix $D^{\cal I}(R)$ of the diquark with spin ${\cal I}$ is
equal to $D_d^0(R^W)=1$ for the
scalar diquark and $D_d^1(R^W)=R^W$ for 
the axial vector diquark.

\section{Form factors and semileptonic decay  rates  }
\label{sec:ff}
In this section we give the general parameterization of semileptonic decay
matrix elements and the expressions for decay rates of heavy baryons
with scalar and axial vector light diquarks.  
\subsection{Heavy baryons with the scalar diquark}
\label{sec:hbsd}
The hadronic matrix elements for the semileptonic decay $\Lambda_Q\to
\Lambda_{Q'}$  are parameterized  in terms of six invariant form factors:
\begin{eqnarray}
  \label{eq:llff}
  \langle \Lambda_{Q'}(v',s')|V^\mu|\Lambda_Q(v,s)\rangle&=& \bar
  u_{\Lambda_{Q'}}(v',s')\Bigl[F_1(w)\gamma^\mu+F_2(w)v^\mu+F_3(w)v'^\mu\Bigl]
u_{\Lambda_Q}(v,s),\cr
 \langle \Lambda_{Q'}(v',s')|A^\mu|\Lambda_Q(v,s)\rangle&=& \bar
  u_{\Lambda_{Q'}}(v',s')\Bigl[G_1(w)\gamma^\mu+G_2(w)v^\mu+G_3(w)v'^\mu\Bigl]
\gamma_5 u_{\Lambda_Q}(v,s),\qquad 
\end{eqnarray}
where   $u_{\Lambda_{Q}}(v,s)$ and
$u_{\Lambda_{Q'}}(v',s')$ are Dirac spinors of the initial and final
baryon with four-velocities $v$ and $v'$, respectively;
$q=M_{\Lambda_{Q'}}v'-M_{\Lambda_Q}v$,  and 
$$w=v\cdot v'=\frac{M_{\Lambda_Q}^2+M_{\Lambda_{Q'}}^2-q^2}
{2M_{\Lambda_Q}M_{\Lambda_{Q'}}}.$$ 

The helicity amplitudes are expressed in terms of these form factors
\cite{kk} as
\begin{eqnarray}
  \label{eq:ha}
  H^{V,A}_{1/2,\, 0}&=&\frac1{\sqrt{q^2}}{\sqrt{2M_{\Lambda_Q}M_{\Lambda_{Q'}}(w\mp 1)}}
[(M_{\Lambda_Q} \pm M_{\Lambda_{Q'}}){\cal F}^{V,A}_1(w) \pm M_{\Lambda_{Q'}}
(w\pm 1){\cal F}^{V,A}_2(w)\cr
&& \pm M_{\Lambda_{Q}} (w\pm 1){\cal F}^{V,A}_3(w)],\cr
 H^{V,A}_{1/2,\, 1}&=&-2\sqrt{M_{\Lambda_Q}M_{\Lambda_{Q'}}(w\mp 1)}
 {\cal F}^{V,A}_1(w),
\end{eqnarray}
where  the upper(lower)  sign corresponds  to $V(A)$ and ${\cal F}^V_i\equiv F_i$,
${\cal F}^A_i\equiv G_i$ ($i=1,2,3$). $H^{V,A}_{\lambda',\,
  \lambda_W}$ are the helicity  
amplitudes for weak transitions induced by vector ($V$) and axial
vector ($A$) currents, where $\lambda'$ and $\lambda_W$ are the
helicities of the final baryon and the virtual $W$-boson, respectively. 
The amplitudes for negative values of the helicities can be obtained
using the relation
$$H^{V,A}_{-\lambda',\,-\lambda_W}=\pm H^{V,A}_{\lambda',\, \lambda_W}.$$
The total helicity amplitude for the
$V-A$ current is then given by
$$H_{\lambda',\, \lambda_W}=H^{V}_{\lambda',\, \lambda_W}
-H^{A}_{\lambda',\, \lambda_W}.$$

The total differential decay rate
\begin{equation}\label{eq:dgtot}
\frac{d\Gamma}{dw}=\frac{d\Gamma_T}{dw}+\frac{d\Gamma_L}{dw}
\end{equation} 
is expressed in terms of the partial rates for transversely ($T$) and
longitudinally ($L$) polarized $W$-bosons
\begin{eqnarray}
  \label{eq:ddrlt}
 \frac{d\Gamma_T}{dw}&=&\frac{G_F^2}{(2\pi)^3} |V_{QQ'}|^2\frac{q^2
   M_{\Lambda_{Q'}}^2\sqrt{w^2-1}}{12M_{\Lambda_Q}} [|H_{1/2,\, 1}|^2+
   |H_{-1/2,\, -1}|^2],\cr
\frac{d\Gamma_L}{dw}&=&\frac{G_F^2}{(2\pi)^3} |V_{QQ'}|^2\frac{q^2
   M_{\Lambda_{Q'}}^2\sqrt{w^2-1}}{12M_{\Lambda_Q}} [|H_{1/2,\, 0}|^2+
   |H_{-1/2,\, 0}|^2],
\end{eqnarray}
where $G_F$ is the Fermi coupling constant and $V_{QQ'}$ is the relevant
CKM matrix element.

The decay products in the semileptonic decay $\Lambda_Q\to
\Lambda_{Q'}(\to \Lambda\pi) +W(\to l\nu)$ are highly polarized. The
polarization of 
the decay products is usually expressed through different asymmetry
parameters \cite{kk} defined  as follows:
\begin{eqnarray}
  \label{eq:da}
 a_T&=&\frac{|H_{1/2,\,1}|^2-|H_{-1/2,\,-1}|^2}{|H_{1/2,\,1}|^2+|H_{-1/2,\,-1}|^2},\cr
a_L&=&\frac{|H_{1/2,\,0}|^2-|H_{-1/2,\,0}|^2}{|H_{1/2,\,0}|^2+|H_{-1/2,\,0}|^2},\cr
\alpha'&=&\frac{|H_{1/2,\,1}|^2-|H_{-1/2,\,-1}|^2}{|H_{1/2,\,1}|^2+|H_{-1/2,\,-1}|^2+
  2(|H_{1/2,\,0}|^2+|H_{-1/2,\,0}|^2)},\cr 
\alpha''&=&\frac{|H_{1/2,\,1}|^2-|H_{-1/2,\,-1}|^2-
  2(|H_{1/2,\,0}|^2+|H_{-1/2,\,0}|^2)}{|H_{1/2,\,1}|^2+|H_{-1/2,\,-1}|^2+ 
  2(|H_{1/2,\,0}|^2+|H_{-1/2,\,0}|^2)},\cr
\gamma&=&\frac{2{\rm Re}(H_{-1/2,\,0}H^*_{1/2,\,1}
  +H_{1/2,\,0}H^*_{-1/2,-\,1})}{|H_{1/2,\,1}|^2+|H_{-1/2,\,-1}|^2+
  |H_{1/2,\,0}|^2+|H_{-1/2,\,0}|^2}. 
\end{eqnarray}
The average values of these asymmetry parameters ($\langle
a_T\rangle$, $\langle a_L\rangle$, $\langle \alpha' \rangle$, $\langle
\alpha'' \rangle$, $\langle \gamma \rangle$) are calculated by
separately integrating the numerators and denominators in
(\ref{eq:da}) over $w$. 
The average longitudinal $\Lambda_{Q'}$ polarization
$\langle P_L\rangle$ can be expressed in terms of $\langle
a_T\rangle$, $\langle a_L\rangle$ as
\begin{equation}
  \label{eq:pl}
\langle P_L\rangle=\frac{ \langle a_T\rangle+R \langle a_L\rangle}
{1+R}, \qquad R=\frac{\Gamma_L}{\Gamma_T}. 
\end{equation}

\subsection{Heavy baryons with the axial vector diquark}
\label{sec:hbavd}
The hadronic matrix elements for the semileptonic decay $\Omega_Q\to
\Omega_{Q'}$  are parameterized  in terms of six invariant form factors
through expressions analogous to Eqs.~(\ref{eq:llff}). Then the
helicity amplitudes and differential decay rates are given by
Eqs.~(\ref{eq:ha}),~(\ref{eq:ddrlt}) with obvious mass replacements.  

The invariant
parameterization for the semileptonic decay $\Omega_Q\to
\Omega^*_{Q'}$ reads:
\begin{eqnarray}
  \label{eq:llffs}
  \langle \Omega^*_{Q'}(v',s')|V^\mu|\Omega_Q(v,s)\rangle= \bar
  u_{\Omega^*_{Q'},\lambda }(v',s')\!\!&\Bigl[&\!\! N_1(w)v^\lambda
\gamma^\mu+N_2(w)v^\lambda v^\mu\cr
&&+N_3(w)v^\lambda v'^\mu+ N_4(w)
g^{\lambda\mu}\Bigl]\gamma_5 u_{\Omega_Q}(v,s),\cr
 \langle \Omega^*_{Q'}(v',s')|A^\mu|\Omega_Q(v,s)\rangle= \bar
  u_{\Omega^*_{Q'},\lambda}(v',s')\!\!&\Bigl[&\!\!K_1(w)v^\lambda\gamma^\mu
+K_2(w)v^\lambda v^\mu\cr
&&+K_3(w)v^\lambda v'^\mu+ K_4(w)g^{\lambda\mu}\Bigl]
u_{\Omega_Q}(v,s),
\end{eqnarray}
where  $u_{\Omega^*_{Q},\mu}$ is the Rarita-Schwinger spinor for
the $\Omega^*_{Q}$, which obeys
$$\not\! v u_{\Omega^*_{Q},\mu}(v,s)=u_{\Omega^*_{Q},\mu}(v,s),
\quad v^\mu u_{\Omega^*_{Q},\mu}(v,s)=\gamma^\mu u_{\Omega^*_{Q},\mu}(v,s)=0.$$

The helicity amplitudes are given by \cite{kk}
\begin{eqnarray}
  \label{eq:haad}
  H^{V,A}_{1/2,\, 0}&=&\mp\frac1{\sqrt{q^2}}\frac{2}{\sqrt3}
\sqrt{M_{\Omega_Q}M_{\Omega^*_{Q'}}(w\mp
  1)}[(M_{\Omega_Q}w-M_{\Omega^*_{Q'}}){\cal N}^{V,A}_4(w)\cr
&&\mp
(M_{\Omega_Q}\mp M_{\Omega^*_{Q'}})(w\pm1){\cal N}^{V,A}_1(w)
+M_{\Omega^*_{Q'}}(w^2-1){\cal N}^{V,A}_2(w)\cr
&&+M_{\Omega_Q}(w^2-1){\cal N}^{V,A}_3(w)],\cr
 H^{V,A}_{1/2,\, 1}&=&\sqrt{\frac23}\sqrt{M_{\Omega_Q}M_{\Omega^*_{Q'}}(w\mp
  1)}[{\cal N}^{V,A}_4(w)-2(w\pm 1){\cal N}^{V,A}_1(w)],\cr
H^{V,A}_{3/2,\, 1}&=&\mp\sqrt{2M_{\Omega_Q}M_{\Omega^*_{Q'}}(w\mp
  1)}{\cal N}^{V,A}_4(w),
\end{eqnarray}
where again the upper(lower)  sign corresponds  to $V(A)$ and ${\cal
  N}^V_i\equiv N_i$, ${\cal N}^A_i\equiv K_i$ ($i=1,2,3$).
The remaining helicity amplitudes  can be obtained
using the relation
$$H^{V,A}_{-\lambda',\,-\lambda_W}=\mp H^{V,A}_{\lambda',\, \lambda_W}.$$
Partial differential decay rates can be represented in the following
form 
\begin{eqnarray}
  \label{eq:darlt}
 \frac{d\Gamma_T}{dw}&=&\frac{G_F^2}{(2\pi)^3} |V_{QQ'}|^2\frac{q^2
   M_{\Omega^*_{Q'}}^2\sqrt{w^2-1}}{12M_{\Omega_Q}} [|H_{1/2,\, 1}|^2+
   |H_{-1/2,\, -1}|^2+|H_{3/2,\, 1}|^2+
   |H_{-3/2,\, -1}|^2],\cr
\frac{d\Gamma_L}{dw}&=&\frac{G_F^2}{(2\pi)^3} |V_{QQ'}|^2\frac{q^2
   M_{\Omega^*_{Q'}}^2\sqrt{w^2-1}}{12M_{\Omega_Q}} [|H_{1/2,\, 0}|^2+
   |H_{-1/2,\, 0}|^2].
\end{eqnarray}
\section{Semileptonic decays of heavy baryons with the scalar diquark}
\label{sec:lld}

In the heavy quark limit $m_Q\to\infty$ ($Q=b,c$) the form factors
(\ref{eq:llff}) 
can be expressed through the single Isgur-Wise function $\zeta(w)$
\cite{iw} 
\begin{eqnarray}
  \label{eq:ffhl}
  F_1(w)&=&G_1(w)=\zeta(w),\cr
F_2(w)&=&F_3(w)=G_2(w)=G_3(w)=0.
\end{eqnarray}
At subleading order of the heavy quark expansion two
additional types of contributions arise \cite{fn}. The first one
parameterizes 
$1/m_Q$ corrections to the HQET current and is proportional to the
product of the parameter $\bar \Lambda=M_{\Lambda_{Q}}-m_Q$, 
which is the difference of
the baryon and heavy quark masses in the infinitely heavy quark limit,
and the leading order Isgur-Wise function $\zeta(w)$. The second one
comes from the 
kinetic energy term in $1/m_Q$ correction to the HQET Lagrangian and
introduces the additional function $\chi(w)$. Therefore the form factors
are given by \cite{fn}
\begin{eqnarray}
  \label{eq:ffso}
  F_1(w)&=& \zeta(w)+\left(\frac{\bar\Lambda}{2m_Q}+
\frac{\bar\Lambda}{2m_{Q'}}\right)\left[2\chi(w)
+ \zeta(w)\right],\cr
 G_1(w)&=& \zeta(w)+\left(\frac{\bar\Lambda}{2m_Q}
+\frac{\bar\Lambda}{2m_{Q'}}\right)\left[2\chi(w)
+\frac{w-1}{w+1} \zeta(w)\right],\cr
F_2(w)&=&G_2(w)=-\frac{\bar\Lambda}{2m_{Q'}}\frac{2}{w+1}\zeta(w),\cr
F_3(w)&=&-G_3(w)=-\frac{\bar\Lambda}{2m_{Q}}\frac{2}{w+1}\zeta(w).
\end{eqnarray}

To calculate these semileptonic decay form factors in our model we
substitute the vertex functions $\Gamma^{(1)}$ (\ref{gamma1}) and
$\Gamma^{(2)}$ (\ref{gamma2}) in the weak current matrix element (\ref{mxet})
between $\Lambda_{Q}$ and $\Lambda_{Q'}$ baryons. It is important to
take into account the relativistic transformation of the baryon wave
functions (\ref{wig}) in this matrix element. The resulting structure
of the decay matrix element is rather complicated, because it is
necessary to integrate both over $d^3p$ and $d^3q$. The $\delta$
function in expression (\ref{gamma1}) for $\Gamma^{(1)}$ permits us to
perform one of these integrations and thus this contribution can be
easily calculated. The calculation of the contribution of the vertex
function  $\Gamma^{(2)}$ (\ref{gamma2}) is more difficult, since
here, instead of a
$\delta$ function, we have a complicated structure,
containing the heavy-quark--light-diquark interaction potential. 
Nevertheless, the
application of the heavy quark $1/m_Q$ expansion considerably
simplifies the
calculation. We carry out this expansion up to the first
order. Then we use the quasipotential equation to perform one of the
integrations in the decay matrix element. The vertex function
$\Gamma^{(1)}$ provides the leading order contribution, while
$\Gamma^{(2)}$ contributes already at the subleading order. The
resulting expressions for the semileptonic decay $\Lambda_Q\to
\Lambda_{Q'}$  form factors up to subleading order in $1/m_Q$ are then
given by  
\begin{eqnarray}
  \label{eq:ffm}
  F_1(w)&=&\zeta(w)+\left(\frac{\bar\Lambda}{2m_Q}
+\frac{\bar\Lambda}{2m_{Q'}}\right)\left[2\chi(w)
+ \zeta(w)\right]\cr
&&+4(1-\varepsilon)(1+\kappa)\left[\frac{\bar\Lambda}{2m_{Q'}}\frac1{w-1}-
\frac{\bar\Lambda}{2m_Q}(w+1)\right]\chi(w),\cr
G_1(w)&=& \zeta(w)+\left(\frac{\bar\Lambda}{2m_Q}
+\frac{\bar\Lambda}{2m_{Q'}}\right)\left[2\chi(w)
+\frac{w-1}{w+1} \zeta(w)\right]\cr
&&-4(1-\varepsilon)(1+\kappa)\frac{\bar\Lambda}{2m_Q}w\chi(w),\cr
F_2(w)&=&-\frac{\bar\Lambda}{2m_{Q'}}\frac{2}{w+1}\zeta(w)\cr
&&-4(1-\varepsilon)(1+\kappa)
\left[\frac{\bar\Lambda}{2m_{Q'}}\frac1{w-1}
+\frac{\bar\Lambda}{2m_Q}w\right]\chi(w),\cr
G_2(w)&=&-\frac{\bar\Lambda}{2m_{Q'}}\frac{2}{w+1}\zeta(w)
-4(1-\varepsilon)(1+\kappa)
\frac{\bar\Lambda}{2m_{Q'}}\frac1{w-1}\chi(w),\cr
F_3(w)&=&-G_3(w)=-\frac{\bar\Lambda}{2m_{Q}}\frac{2}{w+1}\zeta(w)
+4(1-\varepsilon)(1+\kappa)
\frac{\bar\Lambda}{2m_{Q}}\chi(w),
\end{eqnarray}
where the leading order Isgur-Wise function of heavy baryons
\begin{equation}
 \label{eq:iwf}
 \zeta(w)=\lim_{m_Q\to\infty}\int\frac{d^3p}{(2\pi)^3}\Psi_{\Lambda_{Q'}}\!\!
\left({\bf p}+2\epsilon_d(p)\sqrt{\frac{w-1}{w+1}}\ \bf e_\Delta\right)
\Psi_{\Lambda_{Q}}({\bf p}),
\end{equation}
and the subleading  function
\begin{equation}
  \label{eq:eta}
  \chi(w)=-\frac{w-1}{w+1}\lim_{m_Q\to\infty}
\int\frac{d^3p}{(2\pi)^3}\Psi_{\Lambda_{Q'}}\!\!
\left({\bf p}+2\epsilon_d(p)\sqrt{\frac{w-1}{w+1}}\ \bf e_\Delta \right)
\frac{\bar \Lambda-\epsilon_d(p)}{2\bar \Lambda}\Psi_{\Lambda_{Q}}({\bf p}),
\end{equation}
here $\bf e_\Delta={\bf \Delta}/\sqrt{{\bf \Delta}^2}$ is the unit vector in
the direction of   
${\bf \Delta}=M_{\Lambda_{Q'}}{\bf v'}-M_{\Lambda_{Q}}{\bf v}$. It is
important to note that in our model the
expressions for the Isgur-Wise
functions 
$\zeta(w)$ (\ref{eq:iwf}) and $\chi(w)$ (\ref{eq:eta}) are determined
in the whole 
kinematic range accessible in the semileptonic decays in terms of
the overlap integrals of the heavy baryon wave functions, which are
known from the baryon mass spectrum calculations. Therefore we do
not need to make any assumptions about the baryon wave functions
or/and extrapolate our form factors from the single kinematic point,
as it was done in most of previous calculations.

\begin{figure}%[htb]
  \centering
  \includegraphics[width=9cm,angle=-90]{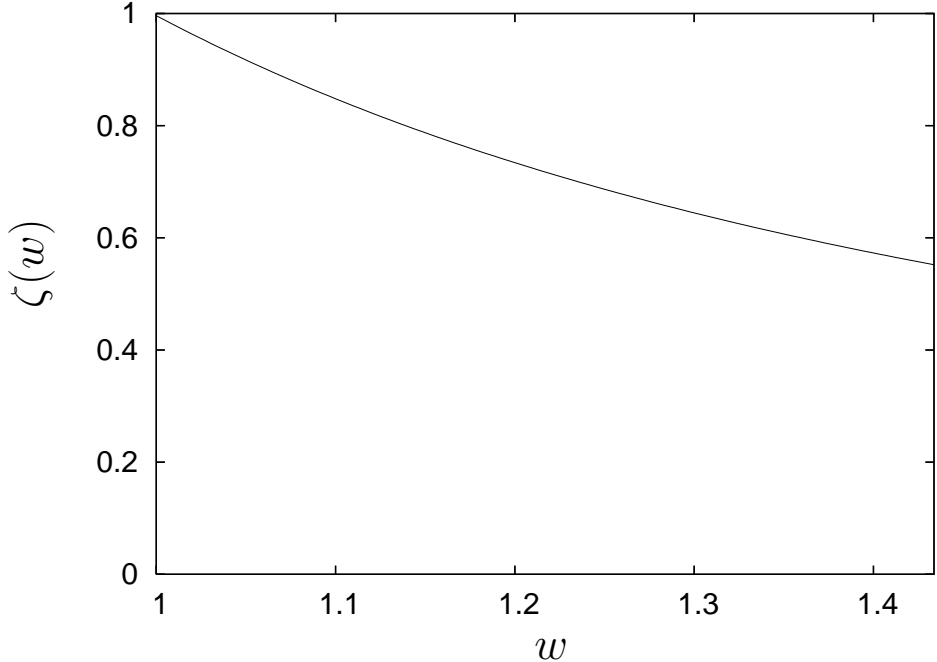}
  \caption{The Isgur-Wise function $\zeta(w)$ for 
the $\Lambda_b\to \Lambda_c e\nu$ semileptonic decay.} 
  \label{fig:xilbc}
\end{figure}

\begin{figure}%[htb]
  \centering
  \includegraphics[width=9cm,angle=-90]{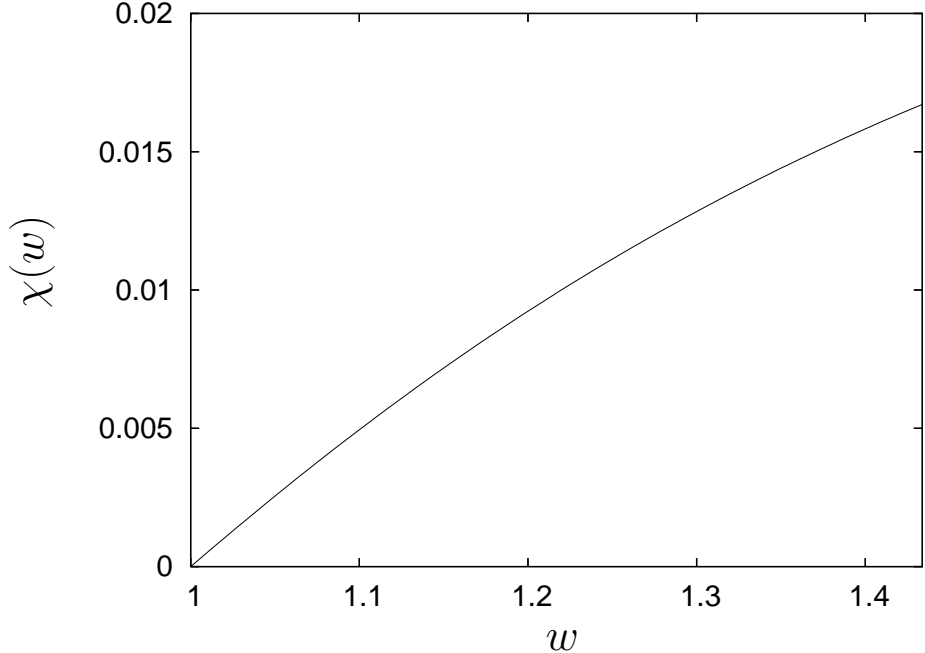}
  \caption{The subleading function $\chi(w)$ for
the $\Lambda_b\to
    \Lambda_c e\nu$ semileptonic decay.} 
  \label{fig:etalbc}
\end{figure}

For $(1-\varepsilon)(1+\kappa)=0$ the HQET results (\ref{eq:ffso}) are
reproduced. This can be achieved either setting $\varepsilon=1$ (pure
scalar confinement) or $\kappa=-1$. In our model we need a
vector
confining contribution (see Sec.~\ref{sec:rqm}) and therefore use the
latter option. This consideration gives us an
additional justification, based on the HQET, 
for fixing  one of the main parameters of the
model $\kappa$.\footnote{It is important to note that the same value of
$\kappa$ is needed 
to get agreement  with the HQET structure of the first order $1/m_Q$
corrections for $B\to D^{(*)} e\nu$ decays \cite{fg}.}    
In the heavy quark limit the wave
functions of the initial $\Psi_{\Lambda_{Q}}$ and final baryon
$\Psi_{\Lambda_{Q'}}$ coincide, and thus the HQET normalization
condition $\zeta(1)=1$ is exactly reproduced. The subleading function
$\chi(w)$ vanishes for $w=1$. These functions, calculated with 
model wave functions for $\Lambda_b$ and $\Lambda_c$ baryons, are
plotted in Figs.~\ref{fig:xilbc},~\ref{fig:etalbc}. The function
$\chi(w)$ is very small in the whole accessible kinematic range, since
it is roughly proportional to the ratio of the heavy baryon binding energy
to the baryon mass.

Near the zero recoil point of the final baryon $w=1$ these
functions can be approximated by 
\begin{eqnarray}
  \label{eq:exp}
  \zeta(w)&=&1-\rho_{\zeta}^2(w-1)+c_{\zeta}(w-1)^2+\cdots,\cr
\chi(w)&=&\rho_\chi^2(w-1)+c_\chi(w-1)^2+\cdots,
\end{eqnarray}
where $\rho_{\zeta}^2=-[d\zeta(w)/dw]_{w=1}$ is the slope and
$2c_{\zeta}=[d^2\zeta(w)/d^2w]_{w=1}$ is the curvature of the Isgur-Wise
functions, which are given in Table~\ref{tab:sl}. The values of
$\zeta(1)$ for transitions between physical
$\Lambda_b$ ($\Xi_b$) and $\Lambda_c$ ($\Xi_c$) baryons are slightly
different (by $\sim 0.5\%$) from the heavy quark limit value 1 due 
to the distinction of the $\Lambda_b$ ($\Xi_b$) and $\Lambda_c$ ($\Xi_c$)
baryon wave functions, calculated for finite values
of the heavy quark masses. 

\begin{table}
  \caption{Parameters of the Isgur-Wise functions for the
    $\Lambda_b\to\Lambda_c e\nu$ and $\Xi_b\to\Xi_c e\nu$ decays.}
  \label{tab:sl}
\begin{ruledtabular}
\begin{tabular}{cccccc}
Decay&$\bar\Lambda$ (GeV) &$\rho_{\zeta}^2$&$c_{\zeta}$&
$\rho_\chi^2$& $c_\chi$ \\
\hline
$\Lambda_b\to\Lambda_c e\nu$& 0.764& 1.70& 2.39 & 0.053
&0.029\\
$\Xi_b\to\Xi_c e\nu$&0.970&2.27& 3.87&0.045&0.036
\end{tabular}
\end{ruledtabular}
\end{table}

\begin{figure}%[htb]
  \centering
  \includegraphics[width=9cm,angle=-90]{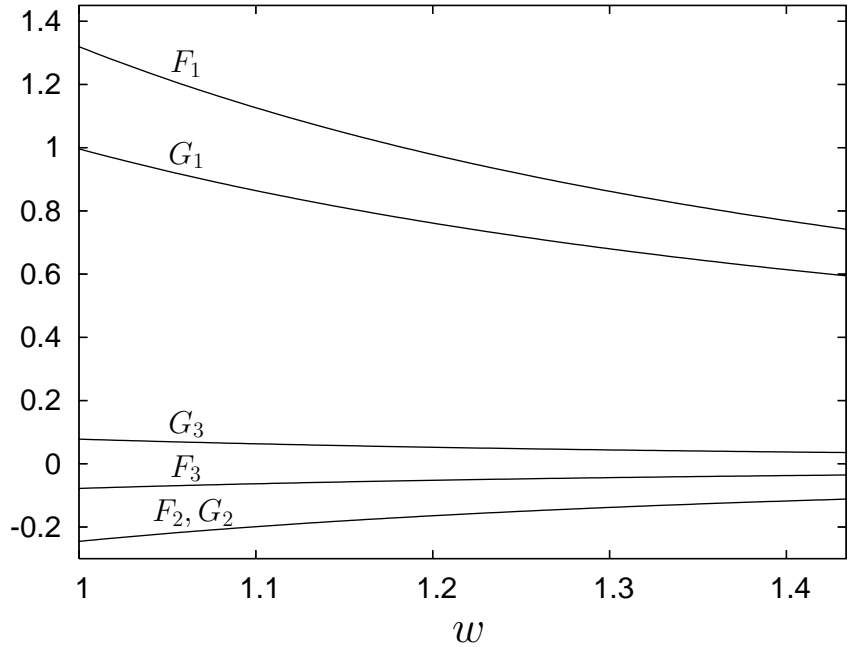}
  \caption{Semileptonic decay form factors for $\Lambda_b\to
    \Lambda_c e\nu$.} 
  \label{fig:fflbc}
\end{figure}

\begin{figure}%[htb]
  \centering
  \includegraphics[width=9cm,angle=-90]{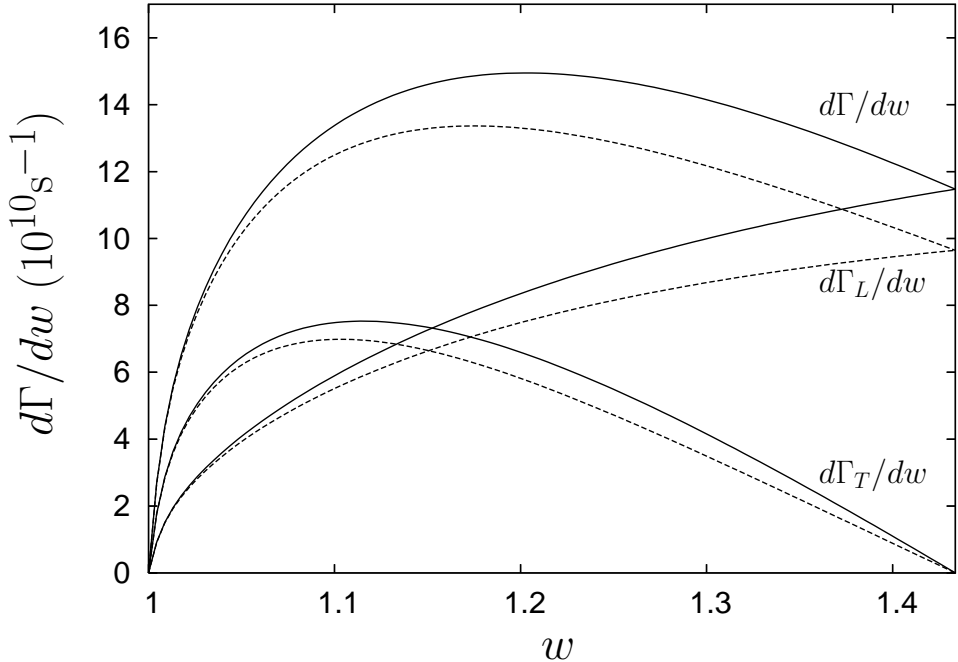}
  \caption{Differential decay rates $d\Gamma/dw$ for 
the $\Lambda_b\to
    \Lambda_c e\nu$ semileptonic decay. Solid lines show decay
  rates including first order $1/m_Q$ corrections. Dashed lines
  correspond to decay rates in the heavy quark limit.} 
  \label{fig:dgammalbc}
\end{figure}

Our model predictions for the form factors $F_i(w)$ and $G_i(w)$
($i=1,2,3$) for the
$\Lambda_b\to\Lambda_c e\nu$ semileptonic decay  are
plotted in Fig.~\ref{fig:fflbc}.  The corresponding differential decay
distributions calculated both with inclusion of
first order heavy quark corrections and in the heavy quark
limit are plotted in Fig.~\ref{fig:dgammalbc}.

The $\Lambda_b\to \Lambda_c$ differential decay rate near zero recoil
\cite{fn}: 
\begin{equation}
  \label{eq:dggg}
  \lim_{w\to 1}\frac1{\sqrt{w^2-1}}\frac{d\Gamma(\Lambda_b\to\Lambda_c
    e\nu)}{dw}=\frac{G_F^2
    |V_{cb}|^2}{4\pi^3}M_{\Lambda_c}^3(M_{\Lambda_b}- M_{\Lambda_c})^2|G_1(1)|^2
\end{equation} 
is governed by the square of the axial current form factor $G_1$,
which near this point has the following expansion
\begin{equation}
  \label{eq:g1exp}
  G_1(w)=1-\hat\rho^2(w-1)+\hat c(w-1)^2+\cdots,
\end{equation}
where in our model with the inclusion of the first order heavy quark
corrections (\ref{eq:ffm}) 
$$\hat\rho^2=1.51,\qquad {\rm and} \qquad \hat c=2.03.$$
This value of the slope parameter of the $\Lambda_b$-baryon decay form
factor is in agreement with the recent experimental value 
obtained by the DELPHI Collaboration
\cite{delphi}
$$\hat\rho^2=2.03\pm0.46^{+0.72}_{-1.00}$$
and lattice QCD \cite{betal} estimate
$$\hat\rho^2=1.1\pm1.0.$$

Our prediction for the branching ratio of the semileptonic decay
$\Lambda_b\to \Lambda_c e\nu$ for   $|V_{cb}|=0.041$ and
$\tau_{\Lambda_b}=1.23\times 10^{-12}$s \cite{pdg} 
$$Br^{\rm theor}(\Lambda_b\to \Lambda_c l\nu)=6.9\%$$
is in agreement with available experimental data
\begin{equation}\label{dcdf}
Br^{\rm exp}(\Lambda_b\to \Lambda_c l\nu)=\left\{
\begin{array}{ll} \left(5.0^{+1.1}_{-0.8}{}^{+1.6}_{-1.2}\right)\% & {\rm
    DELPHI}\ [23]\cr 
\left(8.1\pm 1.2^{+1.1}_{-1.6}\pm 4.3\right)\%  & {\rm CDF}\ [25]
\end{array}\right. \end{equation}
and the PDG branching ratio \cite{pdg}
\begin{equation}\label{dpdg}
Br^{\rm exp}(\Lambda_b\to \Lambda_c l\nu+{\rm
  anything})=(9.1\pm2.1)\%.
\end{equation}

\begin{table}
  \caption{Theoretical predictions for semileptonic decay rates
    $\Gamma$ (in $ 10^{10}{\rm s}^{-1}$) and
    averaged asymmetries for  $\Lambda_b\to\Lambda_c e\nu$ and
    $\Xi_b\to\Xi_c e\nu$ for $|V_{cb}|=0.041$. Branching ratios $Br$
    (in \%) are calculated 
    using experimental mean values \cite{pdg} 
    for the life times  $\tau_{\Lambda_b}=1.23\times 10^{-12}$s and
    $\tau_{\Xi_b}=1.39\times 10^{-12}$s.}
  \label{tab:dr}
\begin{ruledtabular}
\begin{tabular}{cccccccccccc}
Decay&$\Gamma$ &$Br$&$\Gamma_L$  &$\Gamma_T$  &$R$&$\langle
a_T\rangle$ & $\langle a_L\rangle$ & $\langle P_L\rangle$ & $\langle
\alpha'\rangle$&  $\langle \alpha''\rangle$ & $\langle \gamma\rangle$   \\
\hline
\multicolumn{4}{l}{in $m_Q\to\infty$ limit}\\
$\Lambda_b\to\Lambda_c e\nu$& 5.02& 6.2& 3.08& 1.94 & 1.59 & $-0.483$ &
$-0.928$ & $-0.756$ & $-0.116$ & $-0.521$ & 0.562\\
$\Xi_b\to\Xi_c e\nu$&4.64& 6.4&2.79& 1.85&1.51 & $-0.455$ & $-0.920$
&$-0.735$ & $-0.113$ & $-0.503$ & 0.587\\
\multicolumn{4}{l}{with $1/m_Q$ corrections}\\
$\Lambda_b\to\Lambda_c e\nu$& 5.64& 6.9& 3.48& 2.16 & 1.61 & $-0.600$
& $-0.940$ & $-0.810$ & $-0.142$ & $-0.527$ & 0.494 \\
$\Xi_b\to\Xi_c e\nu$&5.29& 7.4&3.21& 2.08&1.54 & $-0.597$ & $-0.935$
&$-0.802$ &$-0.146$ & $-0.510$ & 0.505\\
\end{tabular}
\end{ruledtabular}
\end{table}

Predictions of our model for the semileptonic decay rates
(\ref{eq:ddrlt}) and 
averaged asymmetries (\ref{eq:da}) and (\ref{eq:pl}) for
$\Lambda_b\to\Lambda_c e\nu$ and $\Xi_b\to\Xi_c e\nu$ decays, both in
the heavy quark limit and with inclusion of first order $1/m_Q$
corrections, are given in Table~\ref{tab:dr}. In decay rate
calculations we used for the $\Xi_b$ mass the value from
Table~\ref{tab:bm} and for other 
masses their experimental values \cite{pdg}. Comparing results for the
decay rates  with  and without first order
$1/m_Q$ corrections we see that the inclusion of the subleading terms
leads to a relatively small $\sim 14\%$ increase of the total decay
rates. Therefore, one can 
expect that higher order corrections should be small,
and thus their account
cannot substantially change the leading order predictions.   

\section{Semileptonic decays of heavy baryons with the axial vector
  diquark} 
\label{sec:avd}

In the heavy quark limit $m_Q\to\infty$ the decay matrix element
(\ref{eq:llffs}) is reduced to \cite{iw,bb}  
\begin{equation}
  \label{eq:mhql}
   \langle \Omega^{(*)}_{Q'}(v',s')|\bar h_{v'}^{(Q')}\Gamma  
h_{v}^{(Q)}|\Omega_Q(v,s)\rangle= 
\bar B^{\Omega^{(*)}_{Q'}}_\mu(v',s')\Gamma
B^{\Omega_{Q}}_\nu(v,s)[-g^{\mu\nu}\zeta_1(w)+v^\mu v'{}^\nu \zeta_2(w)],
\end{equation}
where 
\begin{equation}
  \label{eq:bb}
  B^{\Omega_{Q}}_\mu(v,s)=\frac1{\sqrt3}(\gamma_\mu+v_\mu)\gamma_5u_{\Omega_Q}(v,s),
\qquad B^{\Omega^{*}_{Q}}_\mu(v,s)=u_{\Omega^*_Q,\mu}(v,s).
\end{equation}
The  structure of the leading order in $1/m_Q$
corrections, which in the HQET can be parameterized in terms of five
additional functions, can be found in Ref.~\cite{bb}. 

\begin{figure}%[htb]
  \centering
  \includegraphics[width=9cm,angle=-90]{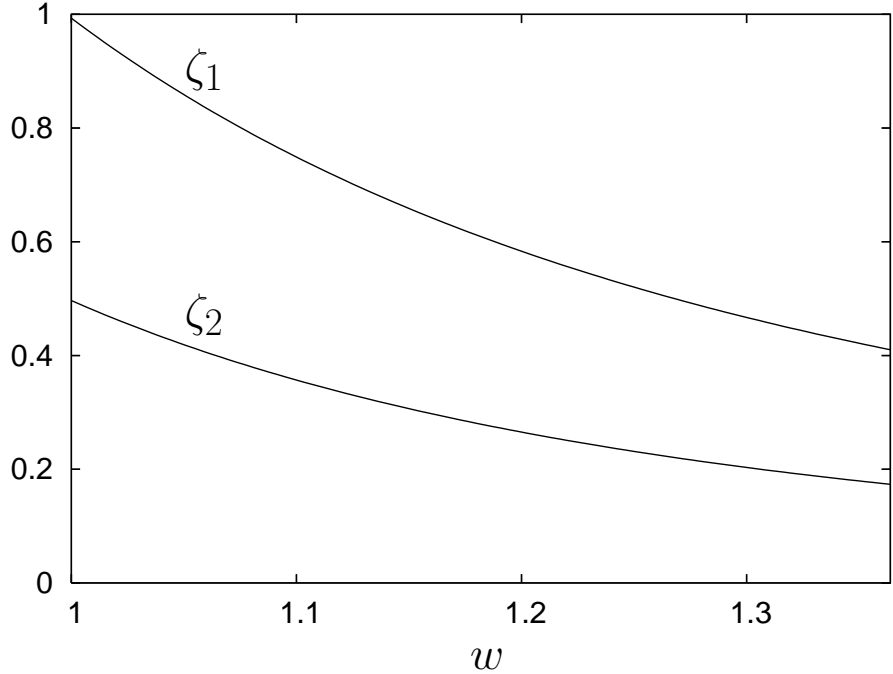}
  \caption{The Isgur-Wise functions $\zeta_1(w)$ and $\zeta_2(w)$ for 
the $\Omega_b\to
    \Omega^{(*)}_c e\nu$ semileptonic decay.} 
  \label{fig:zetao}
\end{figure}

In our model the corresponding semileptonic decay matrix element can
be calculated using the same procedure as in the previous
section. However such calculation is considerably more cumbersome
(especially for the $\Gamma^{(2)}$ contribution),
since now the spectator light diquark has spin equal to 1. Taking into
account that in the $\Lambda_Q$ baryon decays contributions of $1/m_Q$
corrections are rather small, we expect that in the  case of
 $\Omega_Q$ baryon decays such corrections should be also relatively
small. At present no bottom baryons with axial vector diquark
have been observed yet, and, when observed, their semileptonic decays
will be difficult to measure.  Therefore it seems reasonable to limit
our analysis here
to the leading order of the heavy quark expansion. In the
heavy quark limit only the lowest order vertex function 
$\Gamma^{(1)}$ (\ref{gamma1}) contributes to the decay matrix element
(\ref{mxet}).  The resulting expressions for the weak decay matrix
element exactly satisfy the HQET relation (\ref{eq:mhql}) and allow us
to determine the Isgur-Wise functions $\zeta_1(w)$ and $\zeta_2(w)$ in
the whole accessible kinematic range through the overlap integrals of
the baryon wave functions. They are given by 
\begin{eqnarray}
  \label{eq:zeta}
  \zeta_1(w)&=&=\lim_{m_Q\to\infty}
\int\frac{d^3p}{(2\pi)^3}\Psi_{\Omega_{Q'}}\!\!
\left({\bf p}+2\epsilon_d(p)\sqrt{\frac{w-1}{w+1}}\ \bf e_\Delta\right)
\Psi_{\Omega_{Q}}({\bf p}),\\
\zeta_2(w)&=&\frac1{w+1} \zeta_1(w), \label{eq:zeta2}
\end{eqnarray}
where $\bf e_\Delta={\bf \Delta}/\sqrt{{\bf \Delta}^2}$ is the unit
vector in the direction of  
${\bf \Delta}=M_{\Omega_{Q'}}{\bf v'}-M_{\Omega_{Q}}{\bf v}$. The
relation (\ref{eq:zeta2}) follows from the relativistic spin
transformation (\ref{wig}) of the spectator axial vector diquark. 
A similar relation was obtained also in Ref.~\cite{kkp}. The 
Isgur-Wise functions are plotted in Fig.~\ref{fig:zetao}. 
 
Near the zero recoil  point  $w=1$ the Isgur-Wise functions can 
again be approximated by 
\begin{equation}
  \label{eq:expzeta}
  \zeta_{i}(w)=\zeta_i(1)-\rho_{\zeta_i}^2(w-1)+c_{\zeta_i}(w-1)^2+\cdots,
\end{equation}
where $\zeta_1(1)=1$ and $\zeta_2(1)=1/2$; 
$\rho_{\zeta_i}^2=-[d\zeta_i(w)/dw]_{w=1}$ is the slope and
$2c_{\zeta_i}=[d^2\zeta_i/d^2w]_{w=1}$ is the curvature of the Isgur-Wise
functions, which are given in Table~\ref{tab:slzeta}.

\begin{table}
  \caption{Parameters of the Isgur-Wise functions for the
    $\Sigma_b\to\Sigma^{(*)}_c e\nu$,  $\Xi'_b\to\Xi'{}^{(*)}_c e\nu$
    and $\Omega_b\to\Omega^{(*)}_c e\nu$ decays.}
  \label{tab:slzeta}
\begin{ruledtabular}
\begin{tabular}{cccccc}
Decay&$\bar\Lambda$ (GeV)   &$\rho_{\zeta_1}^2$&$c_{\zeta_1}$
&$\rho_{\zeta_2}^2$& $c_{\zeta_2}$ \\
\hline
$\Sigma_b\to\Sigma^{(*)}_c e\nu$& 0.942& 2.17& 3.62& 1.34
&2.44\\
$\Xi'_b\to\Xi'{}^{(*)}_c e\nu$&1.082&2.61& 4.93&
1.55&3.19\\
$\Omega_b\to\Omega^{(*)}_c e\nu$&1.208&2.99& 6.21&
1.74&3.91
\end{tabular}
\end{ruledtabular}
\end{table}

The invariant form factors in the heavy quark limit can be expressed,
using relation (\ref{eq:zeta2}),
in terms of the Isgur-Wise function $\zeta_1(w)$  as follows 
\begin{eqnarray}
  \label{eq:ffo}
  F_1(w)&=&G_1(w)=-\frac13\zeta_1(w),\cr
F_2(w)&=&\frac23\frac2{w+1}\zeta_1(w),\cr
G_2(w)&=&G_3(w)=0;\cr
N_1(w)&=&-N_3(w)=K_3(w)=-\frac1{\sqrt3}\frac2{w+1}\zeta_1(w),\cr
N_4(w)&=&-K_4(w)=-\frac2{\sqrt3}\zeta_1(w),\cr
N_2(w)&=&K_1(w)=K_2(w)=0.
\end{eqnarray}

\begin{figure}%[htb]
  \centering
  \includegraphics[width=9cm,angle=-90]{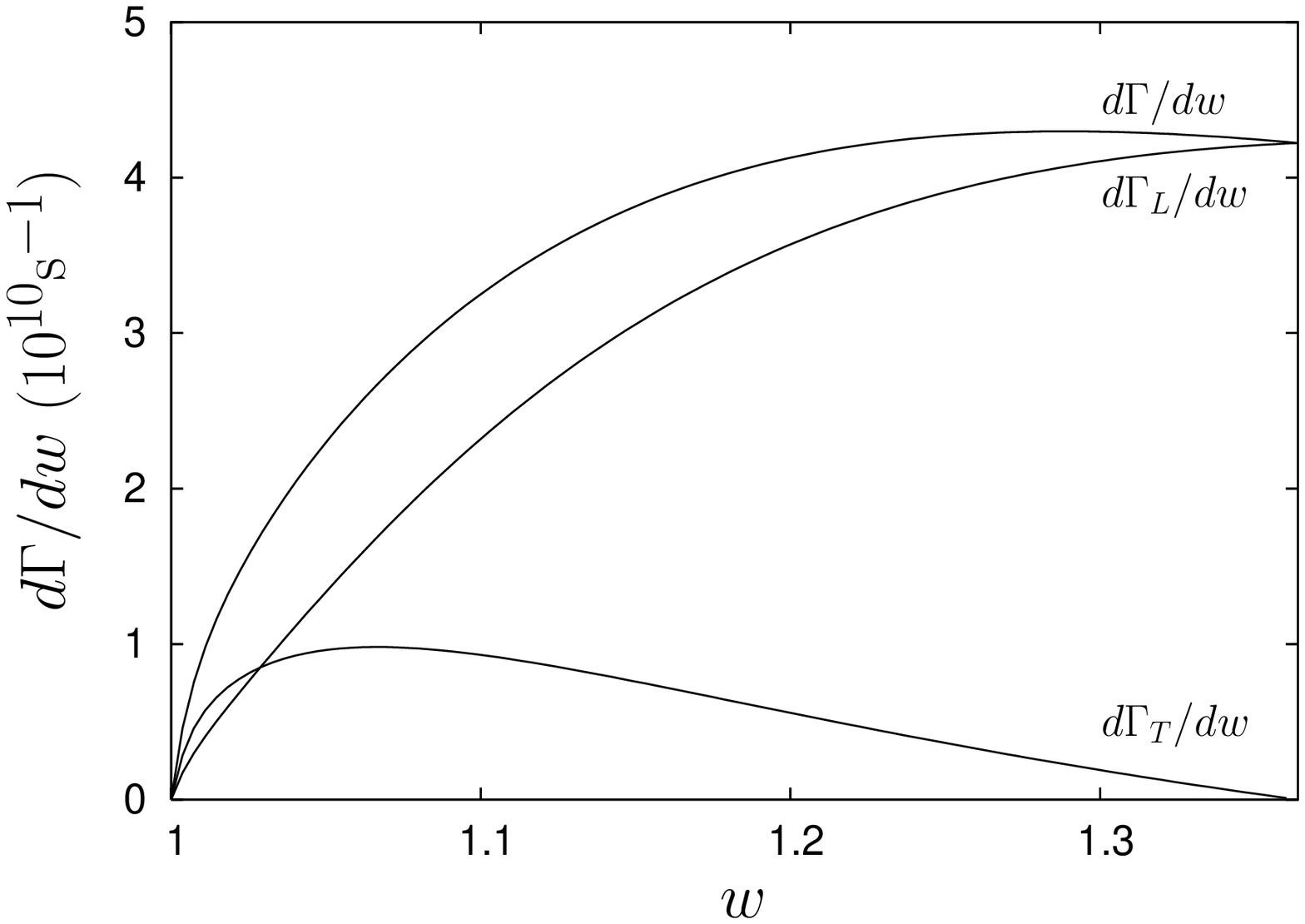}
  \caption{Differential decay rates $d\Gamma/dw$ for 
the $\Omega_b\to
    \Omega_c e\nu$ semileptonic decay. } 
  \label{fig:dgammao}
\end{figure}

\begin{figure}%[htb]
  \centering
  \includegraphics[width=9cm,angle=-90]{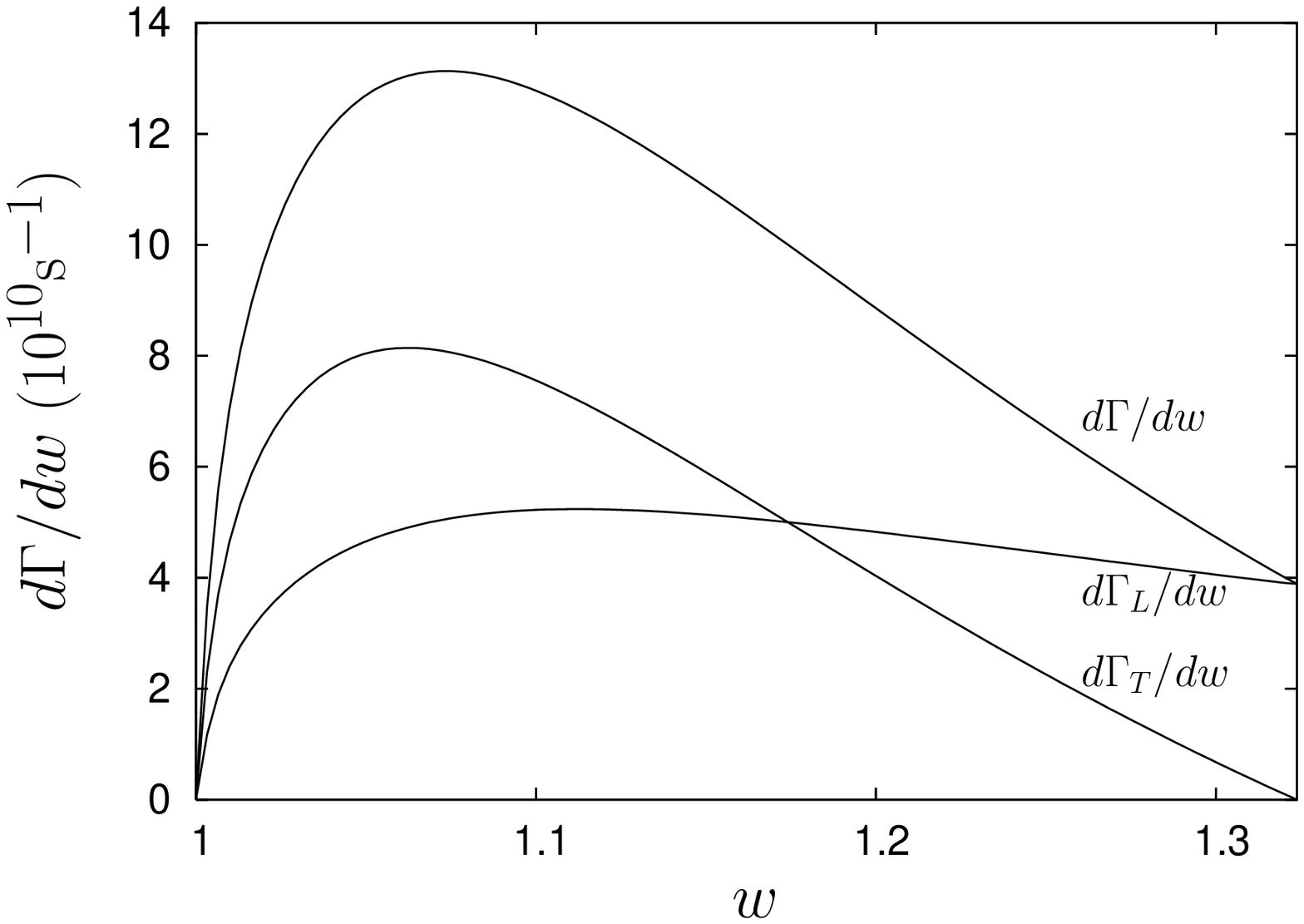}
  \caption{Differential decay rates $d\Gamma/dw$ for 
the $\Omega_b\to \Omega^*_c e\nu$ semileptonic decay. } 
  \label{fig:dgammaos}
\end{figure}

\begin{table}
  \caption{Theoretical predictions for semileptonic decay rates
    $\Gamma$ (in $ 10^{10}{\rm s}^{-1}$) of bottom baryons with the
    axial vector 
  diquark in the heavy quark limit  for $|V_{cb}|=0.041$.}
  \label{tab:drad}
\begin{ruledtabular}
\begin{tabular}{ccccc}
Decay&$\Gamma$ &$\Gamma_L$  &$\Gamma_T$  &$R$  \\
\hline
$\Sigma_b\to\Sigma_c e\nu$& 1.44& 1.23& 0.21& 5.87\\
$\Xi'_b\to\Xi'_c e\nu$&1.34& 1.14&0.20& 5.59 \\
$\Omega_b\to\Omega_c e\nu$& 1.29& 1.09& 0.20& 5.31 \\
$\Sigma_b\to\Sigma_c^* e\nu$& 3.23& 1.61& 1.62& 0.99\\
$\Xi'_b\to\Xi_c^* e\nu$&3.09& 1.52&1.57& 0.97 \\
$\Omega_b\to\Omega_c^* e\nu$& 3.03& 1.48& 1.55& 0.95 
\end{tabular}
\end{ruledtabular}
\end{table}

The differential decay rates $d\Gamma/dw$ for $\Omega_b\to
 \Omega^{(*)}_c e\nu$ semileptonic decays are plotted in
 Figs.~\ref{fig:dgammao} and \ref{fig:dgammaos}. The decay
rates of bottom baryons with the axial vector diquark, calculated in the
heavy quark limit using expressions (\ref{eq:darlt}), are given in
Table~\ref{tab:drad}. For masses of bottom baryons and the
$\Omega^*_c$ we used the values from Table~\ref{tab:bm} and for
other charmed baryons we used experimental mass values \cite{pdg}.

\section{Discussion and conclusions}
\label{sec:dc}

The comparison of our model predictions with other theoretical
calculations \cite{singl,ct,kkp,ilkk,iklr,cs,ahn,hjkl} is given in
Table~\ref{tab:tdr}. In nonrelativistic quark models \cite{singl,ct,kkp}
form factors of the heavy baryon decays are evaluated at the single
kinematic point of zero recoil and then different form factor
parameterizations (pole, dipole) are used for decay rate calculations.   
The relativistic three-quark model \cite{ilkk}, Bethe-Salpeter
model \cite{iklr} and light-front constituent quark model \cite{cs}
assume Gaussian wave functions for heavy baryons. 
The authors of the recent nonrelativistic quark model \cite{ahn} use
for the form factor evaluations 
the set of variational wave functions, obtained from baryon spectra
calculations without employing the quark-diquark
approximation. Finally, Ref.~\cite{hjkl} presents the recent QCD sum rule
prediction. Calculations of Refs.~\cite{kkp,ilkk,iklr} are done in
the heavy quark limit only, while the rest include first order $1/m_Q$
corrections for the decays of $\Lambda$-type baryons. From
Table~\ref{tab:tdr} we see that all theoretical 
models give close predictions for the semileptonic decays of heavy
baryons with scalar diquark ($\Lambda_b\to\Lambda_c e\nu$ and
$\Xi_b\to\Xi_c e\nu$), which are consistent with the available
experimental data (\ref{dcdf}) and (\ref{dpdg}) for the
$\Lambda_b\to\Lambda_c e\nu$ semileptonic decay. The results for
averaged asymmetries of 
these decays (see Table~\ref{tab:dr}) are also close 
in most of the considered approaches. Thus one can conclude that the
precise measurement of the semileptonic $\Lambda_b\to\Lambda_c e\nu$
decay rate will allow an accurate determination of the CKM matrix
element $V_{cb}$ with small theoretical uncertainties.

\begin{table}
  \caption{Comparison of different theoretical predictions for
    semileptonic decay rates $\Gamma$ (in $ 10^{10}{\rm s}^{-1}$) of
    bottom baryons.} 
  \label{tab:tdr}
\begin{ruledtabular}
\begin{tabular}{cccccccccc}
Decay&this work &\cite{singl}& \cite{ct}& \cite{kkp} &\cite{ilkk}
&\cite{iklr}& \cite{cs}&\cite{ahn} &\cite{hjkl}\\
\hline
$\Lambda_b\to\Lambda_c e\nu$& 5.64& 5.9& 5.1& 5.14& 5.39& 6.09&
$5.08\pm 1.3$& 5.82 &$5.4\pm 0.4$\\
$\Xi_b\to\Xi_c e\nu$&5.29&7.2 & 5.3 & 5.21 & 5.27& 6.42& $5.68\pm 1.5$
& 4.98 &\\
$\Sigma_b\to\Sigma_c e\nu$& 1.44& 4.3 & &  & 2.23&1.65& & &\\
$\Xi'_b\to\Xi'_c e\nu$&1.34& & & & & & & &\\
$\Omega_b\to\Omega_c e\nu$& 1.29&5.4 & 2.3&1.52 &1.87&1.81& & &\\
$\Sigma_b\to\Sigma_c^* e\nu$& 3.23& & & &4.56 &3.75& & &\\
$\Xi'_b\to\Xi_c^* e\nu$ &3.09&  & & & & & & &\\
$\Omega_b\to\Omega_c^* e\nu$& 3.03& & & 3.41& 4.01&4.13 & & & 
\end{tabular}
\end{ruledtabular}
\end{table}

All predictions for heavy baryon decays with the axial vector diquark
listed in Table~\ref{tab:tdr} were obtained in the heavy quark
limit. Here the differences between predictions  are larger. The
nonrelativistic quark model \cite{singl} gives for these decay rates
values more than two times larger than other estimates. Our model
values for these decay rates are the lowest ones.
Among the relativistic quark models
the closest to our predictions is given in
\cite{iklr}. Unfortunately, it will be difficult to measure such
decays experimentally. Only  $\Omega_b$ (which has not been
observed yet) will decay predominantly weakly and thus 
has sizable semileptonic branching fractions, since a
scalar $ss$ diquark is forbidden by the
Pauli principle. All other baryons with the axial vector diquark will
decay predominantly strongly or electromagnetically and thus their
weak branching ratios will be very small.

In summary, in this paper we calculated the semileptonic decay rates
of heavy baryons in the framework of the relativistic quark
model. Heavy baryons were considered in 
the heavy-quark--light-diquark approximation. The baryon wave
functions were obtained previously in the process of the heavy baryon
mass spectrum calculations. In our approach the spectator diquark is
not treated as a point-like object. The relatively large diquark size
is taken into account by calculating the diquark-gluon 
form factor as the overlap integral of the diquark wave functions. The
matrix element of the weak current between baryon states was
considered using the quasipotential approach. The relativistic
transformation of the baryon wave functions from the rest reference
frame to the moving one as well as the negative energy contributions
to the decay matrix elements were explicitly taken into account. To
simplify calculations 
and in order to compare with model-independent predictions of HQET the
heavy quark expansion was applied up to subleading order for heavy
baryon decays with a
scalar light diquark. It was shown that all HQET relations in the
leading and subleading order are exactly satisfied in our model if 
the long-range chromomagnetic interaction vanishes ($\kappa=-1$) in
accord with our previous analysis of heavy meson decays. The leading
and subleading Isgur-Wise functions were determined in terms of
the overlap integrals of baryon wave functions. It was found by
explicit calculation that the
additional subleading function $\chi(w)$, arising from the
kinetic energy
term in the HQET Lagrangian, is negligibly small in the whole kinematic
range. Decay rates as well as different averaged asymmetries both with
and without $1/m_Q$ corrections were 
calculated. Moreover, it was shown that the subleading terms in the
heavy quark
expansion modify the results for decay rates by $\sim 14\%$. Thus one
can expect that the influence of higher order corrections should be
small.  

The decays of heavy baryons with the axial vector diquark were
considered in the heavy quark limit. All HQET relations are exactly
satisfied in our model. The corresponding 
Isgur-Wise functions were
determined in terms of the overlap integrals of the baryon wave
functions. It was found that the 
relativistic transformation of the axial vector diquark spin leads to
the relation (\ref{eq:zeta2}) between baryon Isgur-Wise functions
$\zeta_1(w)$ and $\zeta_2(w)$. 

The calculated decay rates of heavy
baryons were compared with the results of other theoretical approaches
and available experimental data. One of the main advantages of our
model is that it allows one to calculate consistently the heavy baryon
wave functions from the 
consideration of the spectroscopy and then determine through
the wave function overlap integrals the
baryonic Isgur-Wise functions in the whole kinematic range accessible
in semileptonic decays. Thus we do not need to make any assumptions
about the form of the baryon wave functions or/and extrapolate the
form factors from one point to the whole kinematic region using
some ad hoc ansatz. No additional free parameters were introduced in our
calculations. As it was pointed out above, we also consistently 
include relativistic effects. All this makes the presented results
sufficiently accurate and reliable.

\acknowledgments
The authors are grateful to M. A. Ivanov, V. E. Lyubovitskij,
M. M\"uller-Preussker and V. I. Savrin  
for support and useful discussions.  Two of us
(R.N.F. and V.O.G.)  were supported in part by the {\it Deutsche
Forschungsgemeinschaft} under contract Eb 139/2-3 and by the {\it Russian
Foundation for Basic Research} under Grant No.05-02-16243.

\end{document}